\documentclass{aa}  

\usepackage{graphicx}
\usepackage{txfonts}
\begin{document} 


   \title{Compact Disks}

   \subtitle{An explanation to faint CO emission in Lupus disks}

   \author{A. Miotello
          \inst{1}
           \and
           G. Rosotti
          \inst{2}$^,$\inst{3}
           \and
           M. Ansdell
          \inst{4}
          S. Facchini
          \inst{1}
           \and
           C. F. Manara
          \inst{1}
          \and
          J. P. Williams
          \inst{5}
          \and
          S. Bruderer}

   \institute{European Southern Observatory, Karl-Schwarzschild-Str 2, 85748 Garching, Germany
 \and
 Leiden Observatory, Leiden University, P.O. Box 9513, NL-2300 RA Leiden, the Netherlands
 \and
 School of Physics and Astronomy, University of Leicester, Leicester LE1 7RH, UK
 \and
 NASA Headquarters, 300 E Street SW, Washington, DC 20546, USA
 \and
 Institute for Astronomy, University of Hawai‘i at Manoa, Honolulu, HI, USA
 }

   \date{Received February 12, 2021, accepted April 27, 2021}
 
  \abstract
   {ALMA disk surveys have shown that a large fraction of observed protoplanetary disks in nearby Star-Forming Regions (SFRs) are fainter than expected in CO isotopologue emission. Disks not detected in $^{13}$CO line emission are also faint and often unresolved in the continuum emission at an angular resolution of around 0.2 arcseconds.}
   {Focusing on the Lupus SFR, the aim of this work is to investigate whether this population comprises radially extended and low mass disks - as commonly assumed so far - or if it is of intrinsically radially compact disks, an interpretation that we propose in this paper. The latter scenario was already proposed for individual sources or small samples of disks, while this work targets a large population of disks in a single young SFR for which statistical arguments can be made. }
   {A new grid of physical-chemical models of compact disks has been run with the physical-chemical code DALI in order to cover a region of the parameter space that had not been explored before with this code. Such models have been compared with $^{12}$CO and $^{13}$CO ALMA observations of faint disks in the Lupus SFR. The  simulated  integrated  continuum  and  CO  isotopologue fluxes  of  the  new  grid  of  compact  models  are  reported. }
   {Lupus disks that are not detected in $^{13}$CO emission and with faint  or  undetected $^{12}$CO  emission are consistent with compact disk models. For disks with a limited radial extent, the emission of CO isotopologues is mostly optically thick and it scales with the surface area: i.e., it is fainter for smaller objects.  The fraction of compact disks is potentially between roughly 50\% and 60\% of the entire Lupus sample. Deeper observations of $^{12}$CO and $^{13}$CO at a moderate angular resolution will allow us to distinguish whether faint disks are intrinsically compact, or if they are extended but faint, without the need of resolving them. If the fainter end of the disk population observed by ALMA disk surveys is consistent with such objects being very compact, this will either create a tension with viscous spreading or require MHD winds or external processes to truncate the disks.}{}


   \maketitle
%

\section{Introduction}

Thanks to its exquisite angular resolution and unprecedented sensitivity, the Atacama Large Millimeter/submillimeter Array (ALMA) has revolutionised the field of star and planet formation. Together with the very popular high angular resolution images \citep[see e.g.,][]{HLTau,Andrews2018}, ALMA has also significantly enhanced the disk sample size by surveying disks at moderate resolution in many different nearby Star-Forming Regions (SFRs). 
Both dust and gas components have been traced through sub-mm continuum and CO isotopologues rotational line emission in the $\sim1$ Myr-old to $\sim10$ Myr-old Lupus, Chamaeleon I, Orion Nebula Cluster, Ophiuchus, IC348, Taurus, and Corona Australis, $\sigma$-Orionis, $\lambda$-Orionis, and Upper Scorpius regions \citep{Ansdell2016,Pascucci2016,Eisner2016,Cieza2019,Long2018,Cazzoletti2019,Ansdell2017,Barenfeld2016,Ansdell2020}. Most of the disks in the SFRs targeted by ALMA have also been observed in the optical with spectroscopy, in order to constrain their stellar properties and mass accretion rates \citep[see e.g.,][]{HerczegHillenbrand2014,Alcala2017,Manara2017,Manara2020}. 

One of the main results of these surveys, unfortunately carried out with short integration times, is that the dust continuum and CO isotopologues emission is fainter than expected leading to measurements of low dust and gas masses \citep{Ansdell2016,Pascucci2016,Long2017,Miotello2017,Manara2018}. Another peculiar aspect of the surveyed disks, is that the fainter part of the disk population often shows compact unresolved continuum emission and is not detected in CO isotopologues \citep[see e.g.,][]{Long2018,Barenfeld2016,Pietu2014}. Whether the observations only reveal the tip of a faint extended emission, or if these disks are intrinsically compact is still not constrained by available data. It is however critical to distinguish between these opposite scenarios, because of the implications on disk evolution. Viscous evolution would in fact predict large gaseous disks, and, in contrast, small outer radii could be explained by MHD winds or external processes that truncate the disks \citep[see e.g.,][Zagaria et al., in prep.]{Clarke1991,Clarke2007,Vinke2015,Rosotti2018,Lesur2020,Sellek2020,Trapman2020}.

Disks around binary stars represent a category of sources that are expected to have a smaller radial extent, due to the interaction between the two disks. Some recent works have focused on the continuum emission of disks around binary systems and, as predicted by theory, they have shown that such disks extend to smaller radii than the population of disks around single stars \citep{Manara2019,Zurlo2020}. 

The idea that disks with faint CO fluxes may be radially compact is not new. \cite{Barenfeld2016} propose that an explanation for the lack of CO detections in approximately half of the disks with detected continuum emission is that CO is optically thick but has a compact emitting area ($<$40 au). A similar results is found with IRAM Plateau de Bure observations of T Tauri disks by \cite{Pietu2014}, which showed that faint continuum and CO emission in disks often seems to be associated with more compact disks that still have high surface densities in their inner regions. \cite{Pietu2014} also argue that this type of sources could represent up to 25$\%$ of the whole disk population. Furthermore, \cite{Hendler2017} show that the unexpectedly faint [OI] 63 $\mu$m emission of Very Low Mass Stars (VLMSs) observed with the \textit{Herschel} Space Observatory PACS spectrometer is likely indicative of smaller disk sizes than previously thought. Also, source-specific models based on CO upper limits also lead to the conclusion that some disks need to be compact in size, in order to explain their CO non-detections \citep{Woitke2011,Boneberg2018}. Finally, a recent CN study carried out in the entire Lupus sample has shown that for many of the targeted disks, that also show faint CO fluxes, the critical radius $R_{\rm c}$ must be small, even less than 15 au, in order to reproduce the observed low CN fluxes \citep{vanTerwisga2019}.

In contrast with these findings, the physical-chemical disk models run with DALI that were employed to interpret the observations from the Lupus and Chameleon disk surveys were originally tailored to larger and brighter disks \citep{Miotello2016,Miotello2017,Long2017}. A set of more representative DALI models for the fainter and, possibly, more compact disks was missing but needed for a better understanding of the existing population of fainter disks, as also recently noted by \cite{Trapman2021}, and it is presented in Sec \ref{sec:models}. The simulated fluxes are compared with observations of the faint disks in the Lupus SFR - 63 out of 99 sources, presented in Sec. \ref{sec:alma} - and the fraction of potentially compact disk is quantified and discussed in Sec. \ref{Sec:results} and \ref{sec:discussion}. Finally, the simulated integrated continuum and CO isotopologue fluxes of the new grid of compact disk models are reported in Appendix \ref{ancillary} as a new instrument for the interpretation of current and future ALMA observations of disks with faint CO emission.

%
              

\section{ALMA Observations}
\label{sec:alma}
%
For this work we use ALMA Band 6 observations of the continuum and CO isotopologue emission of disks in the Lupus SFR \citep[see][for more details]{Ansdell2018}. More specifically, we focus on the sources that have not been detected in $^{13}$CO $(J=2-1)$ emission and whose $^{12}$CO $(J=2-1)$ luminosity is smaller than $2.5\times 10^{18}$ mJy km s$^{-1}$ pc$^2$. Applying this cut in $^{12}$CO $(J=2-1)$ luminosity, we exclude the brighter and resolved disks where CO outer radii were measured by \cite{Ansdell2018}. Their CO isotopologue fluxes can in fact be explained by physical-chemical models of viscously evolving disks \citep{Trapman2020}. Finally, our sample of faint Lupus disks is composed of 63 disks, out of which only 10 have $^{12}$CO $(J=2-1)$ detections (see Table \ref{table:Lupus}).


The integrated $^{12}$CO and $^{13}$CO ($J=2-1$) line luminosity are presented in Fig. \ref{data_model_Miotello2016} by black squares, where 3$\sigma$ $^{13}$CO and $^{12}$CO upper limits, calculated as three times the rms using an aperture equal to the size of the
typical beam (i.e.$\sim 0.21" - 0.25"$, equivalent to $\sim 30 - 40$ au at 160 pc) which assumes that the disk is only emitting within the beam, are shown by black and gray arrows respectively \citep{Ansdell2018}. Previously unpublished $^{12}$CO fluxes, together with the  $^{13}$CO and $^{12}$CO upper limits of the selected sample of Lupus disks are reported in Table \ref{table:Lupus}.
Some level of cloud absorption affects a few of the Lupus sources considered in this work \citep[see $^{12}$CO spectra in Fig. 11 of][]{Ansdell2018}, whose line luminosities are shown by the empty sqaures in Fig. \ref{data_model_Miotello2016}. Finally, $^{12}$CO and $^{13}$CO non detections are shown  as  upper  limits, calculated as three times the rms, by  grey  arrows. The line luminosities\footnote{Luminosities are calculated following \cite{Williams2014}: $L = 4\pi d^2 F$, where $d$ is the distance of the source and $F$ is its measured spatially integrated flux.} are calculated using the distance of each single object measured by Gaia DR2 \citep{GaiaDR2,Alcala2019}.

The stellar luminosity $L_{\star}$ and stellar mass $M_{\star}$, obtained using the evolutionary track by \cite{Baraffe2015}, of the selected sources are reported in Table \ref{table:Lupus} \citep{Alcala2017}. Many of these sources can be classified as Very Low Mass Stars (VLMSs), having $M_{\star}\lesssim 0.3 M_{\odot}$ \citep{Liebert1987}. Almost all sources that are detected in $^{12}$CO are instead T Tauri-like stars.  \\

\section{Models}
\label{sec:models}

Inspired by the observations presented in Sec. \ref{sec:alma}, we have designed a grid of compact disk physical-chemical models. We use the code DALI \citep[Dust And LInes,][]{Bruderer2012} with a similar setup as in \cite{Miotello2016}. The disk surface density distribution is parametrized by a power-law function, following the prescription proposed by \cite{Andrews2011}: 
\begin{equation}
    \Sigma_{\rm gas}=\Sigma_{\rm c}\left(\frac{R}{R_{\rm c}}\right)^{-\gamma} \exp \left[-\left(\frac{R}{R_{\rm c}}\right)^{2-\gamma}\right],
\end{equation}
where $R_{\rm c}$ is the so-called critical radius. In the large grid of T Tauri-like disk models presented by \cite{Miotello2016}, $R_{\rm c}$ was set to 30, 60 and 200 au, and the power-law index $\gamma$ to 0.8, 1, and 1.5, resulting in disks with non-negligible surface density up to several hundreds of au (see right panels of Fig. \ref{2D}). Such models are not representative of compact disks such as those considered in this work, which show unresolved or marginally resolved continuum emission at a resolution of 36 au (18 au in radius). 

For this study the disk radial extent has been drastically reduced by setting $R_{\rm c}$ to 0.5, 1, 2, 5, and 15 au, and $\gamma$ to 0.5, 1.0, and 1.5. The other disk parameters are also listed below for completeness: disk mass $M_{\rm disk}=10^{-5},10^{-4},10^{-3},10^{-2} M_{\odot}$; scale height $h=0.1$; flaring angle $\psi=0.1$; large-over-small grains mass fraction is $f_{\rm large} = 0.9$, settling parameter $\chi = 0.2$. Two sets of models have been run to cover different stellar parameters. First, T Tauri-like disk models have been run, where the stellar spectrum is composed of a black body with a temperature $T_{\rm eff}=4000$ K and a UV excess which mimics a mass accretion rate of $10^{-8}M_{\odot}\rm yr^{-1}$, the stellar luminosity and mass are set to $L_{\star}=1 L_{\odot}$ as in \cite{Miotello2016}. Since most of the fainter disks observed with ALMA orbit low-mass and low-luminosity young stars, the second set of models uses a synthetic stellar spectrum more representative of the observed stellar parameters. More precisely, the spectrum is composed of a black body with a temperature $T_{\rm eff}=3400$ K and a UV excess which mimics a mass accretion rate of $10^{-9}M_{\odot}\rm yr^{-1}$. The stellar luminosity and mass are set to $L_{\star}=0.16 L_{\odot}$ and $M_{\star}=0.26 M_{\odot}$. From now on, we will refer to these as VLMS-like disk models. We also account for the interstellar UV radiation field \citep{Draine78} and the cosmic microwave background as external sources of radiation. We also consider cosmic rays, for which a rate of $\zeta_{\rm CR} = 5 \times 10^{17}$ s$^{-1}$ is adopted, as a source of ionization \citep[see e.g.,][and references therein]{Bosman2018,Trapman2021}.


\begin{table}[h]
\caption{Parameters of the disk models}             
\label{table:1}      
\centering                          
\begin{tabular}{l l}        
\hline\hline                 
Parameter & Range \\    
\hline                        
 \emph{Chemistry}&\\
   Chemical age & 1 Myr\\
   Chemical network & \cite{Miotello2016}\\
   \hline  
 \emph{Physical Structure}&\\
   $\gamma$ & 0.5, 1, 1.5 \\      
   $\psi$ & 0.1 \\
   $h_{\rm c}$ & 0.1 rad \\
   $R_{\rm c}$ & 0.5, 1, 2, 5, 15 au \\
   $M_{\rm disk}$ & $10^{-5}$, $10^{-4}$, $10^{-3}$, $10^{-2} M_{\odot}$\\
   gas-to-dust ratio& 100 \\
   $f_{\rm large}$ & 0.9\\
   $\chi$ & 0.2 \\
   $i$ & $10^{\rm o}$, 30$^{\rm o}$, 60$^{\rm o}$\\
   \hline  
 \emph{Stellar Spectrum}&\\   
   VLMSs\tablefootmark{**} & $T_{\rm eff}=3400$ K, $L_{\rm bol}=0.26 L_{\odot}$ \\
   T Tauri\tablefootmark{**} & $T_{\rm eff}=4000$ K, $L_{\rm bol}=1 L_{\odot}$ \\
\hline                                   
\end{tabular}
\tablefoot{** FUV excess added, see text.\\}
\end{table}

\section{Results}
\label{Sec:results}

\begin{figure}
   \centering
   \includegraphics[width=9cm]{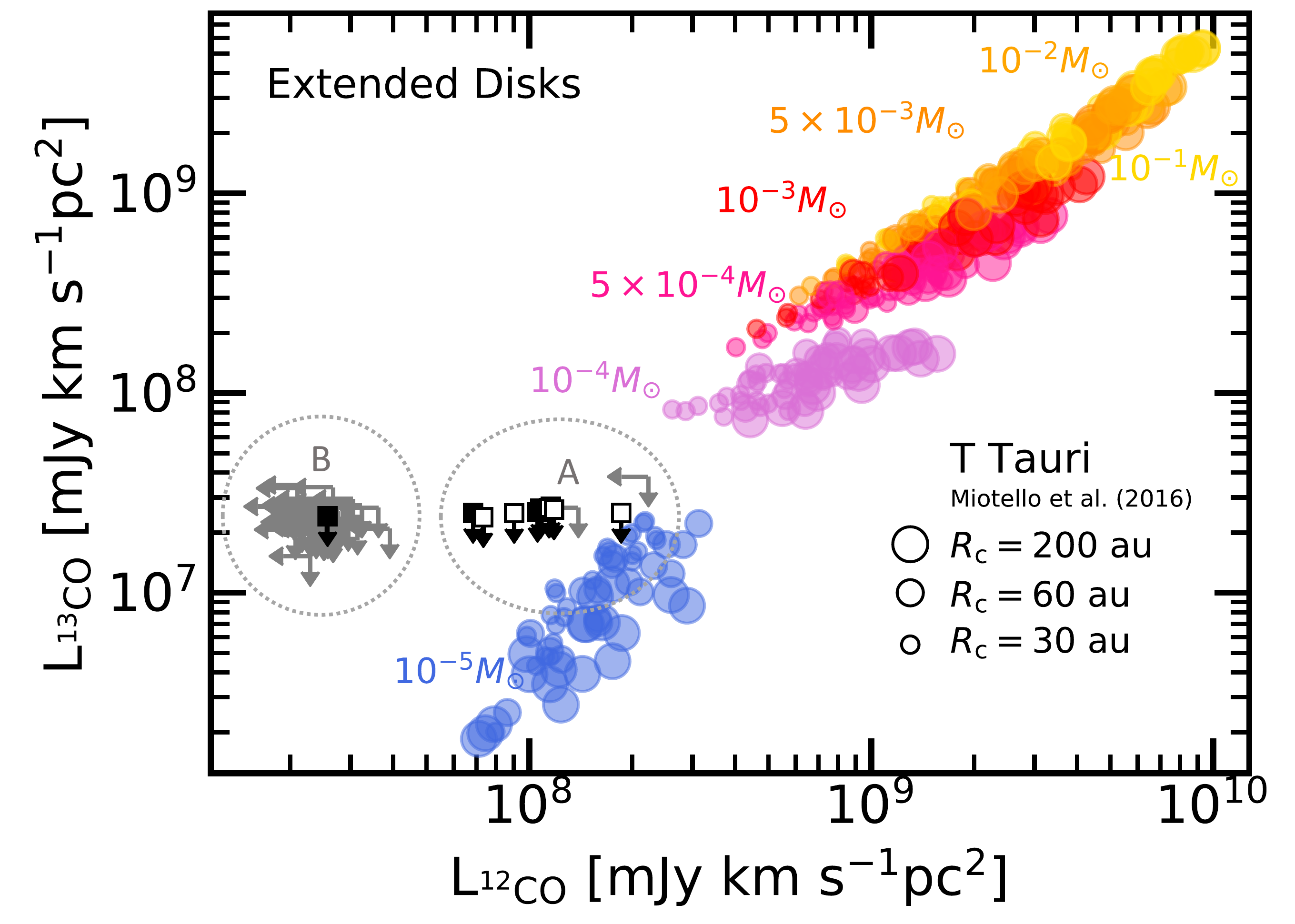}
      \caption{Lupus $^{12}$CO and $^{13}$CO ($J=2-1$) line luminosity are presented with black squares (empty squares if the $^{12}$CO line is partially absorbed by the cloud),where the $^{13}$CO non detections are shown as upper limits by the black arrows and the $^{12}$CO (and $^{13}$CO) non detections are shown as upper limits by the gray arrows. Note that the 3$\sigma$ upper limits are calculated using an aperture equal to the size of the
typical beam \citep[i.e. $\sim 0.21-0.25"$, equivalent to $\sim 30 - 40$ au at 160 pc, see ][]{Ansdell2018}. DALI model results from \cite{Miotello2016} are shown with filled circles, color-coded by disk mass. Different symbol sizes represent different values of the critical radius $R_{\rm c}$.}
         \label{data_model_Miotello2016}
   \end{figure}

Extended disk models previously run with DALI \citep{Miotello2016} with ISM-like volatile C and O abundances may not always be able to simultaneously reproduce $^{12}$CO and $^{13}$CO emission of the observed CO-faint disks in Lupus. This is shown in Fig. \ref{data_model_Miotello2016}, where such models are presented with colored circles and the observations are reported in black and grey. 

The CO-faint disks in Lupus can be divided in two sub-samples, highlighted with two ellipses in Fig. \ref{data_model_Miotello2016}. The group A is composed of 11 disks, mostly detected in $^{12}$CO, with line luminosity $L_{\rm ^{12}CO} \gtrsim 6 \times 10^7$ mJy km s$^{-1}$ pc$^{2}$. Two objects in this group are instead not detected either in $^{12}$CO or in $^{13}$CO. Disks in group A are consistent with the $10^{-5} M_{\odot}$ extended disk models, shown by the blue symbols in Fig. \ref{data_model_Miotello2016}. 

A second set of 52 disks, group B, is composed of disks that are not detected either in $^{13}$CO or in $^{12}$CO , shown by the gray arrows (except for one $^{12}$CO detection, shown by the black square). These are more extreme cases that are not compatible with any of the models by \cite{Miotello2016}.  Accordingly, \cite{Miotello2017} needed to claim high levels of volatile carbon and oxygen depletion as a solution for reaching fainter CO fluxes and match the observed line luminosity.

\begin{figure}
   \centering
   \includegraphics[width=9cm]{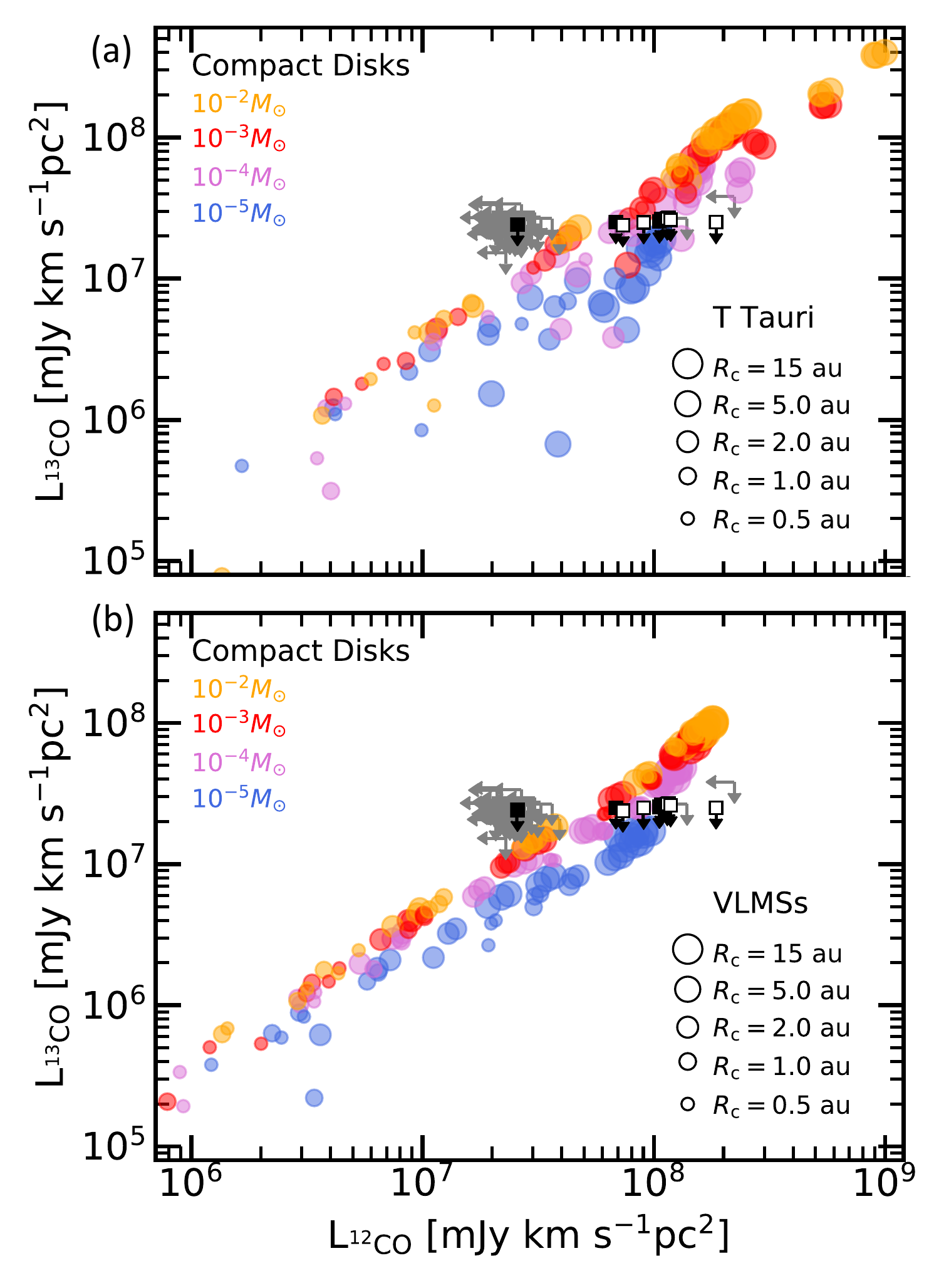}
      \caption{Lupus $^{12}$CO and $^{13}$CO ($J=2-1$) line luminosity are presented with black squares (empty squares if the $^{12}$CO line is partially absorbed by the cloud), where the $^{13}$CO non detections are shown as upper limits by the black arrows and the $^{12}$CO (and $^{13}$CO) non detections are shown as upper limits by the grey arrows. DALI model results for  T Tauri disk models are shown in panel (a), and those for VLMS disk models in panel (b) with filled circles, color-coded by disk mass. Different symbol sizes represent different values of the critical radius $R_{\rm c}$. Notice the different scale from Fig. \ref{data_model_Miotello2016}.}
         \label{data_model}
   \end{figure}

In Fig. \ref{data_model} the results of our new grid of compact disk models are shown with coloured circles in comparison with the observations of the sub-sample of Lupus disks studied in this work. Disk masses are color coded, while different values of $R_{\rm c}$ are shown by different symbol sizes. Compact disk models, with $R_{\rm c}$ smaller than 15 au, produce integrated $^{12}$CO and $^{13}$CO line luminosities that are compatible with Lupus disks in group B, and most sources in group A. Overall, CO luminosities simulated with T Tauri (panel a) and  with VLMSs models (panel b) are not very different, with the first being however higher than the latter. This behaviour is caused by the fact that $^{12}$CO emission is substantially optically thick as the column density reaches very high values in compact disks. Under the optically thick approximation, the intensity of the emission scales directly with the temperature of the emitting material, and it is therefore expected that T Tauri-like disk models reach higher luminosity than VLMS-like disk models with same disk parameters. Such effect is also seen for $^{13}$CO integrated line luminosities, that are generally higher for T Tauri-like disk models. Even $^{13}$CO emission is mostly optically thick in compact disk models: in fact $^{12}$CO and $^{13}$CO integrated line luminosities scale almost linearly as shown in Fig. \ref{data_model}. \cite{Miotello2016} are able to fit integrated line luminosity to simple formulae: in extended disk models $^{13}$CO emission scales directly with disk mass for $M_{\rm disk}<2\times10^{-4}M_{\odot}$ and it can be used as disk mass tracer. On the contrary, for compact disk models $^{13}$CO does not scale linearly with mass, even for the lower mass disk models. This is in line with the findings of \cite{Boneberg2018} and \cite{Greenwood2017}, who, using thermochemical modelling of Brown Dwarfs (BDs) disks, show that CO observations of compact disks are insensitive to disk mass due to high optical depths. One would therefore need optically thinner tracers, such as C$^{18}$O, to trace disk masses. Also for this new grid of models, $^{13}$CO and C$^{18}$O line luminosities can be fitted to logarithmic function, as reported in more detail in Appendix \ref{L-mass}. 

Another consequence of the optically thick approximation is that luminosity directly scales with the surface area of the emitting material, i.e, linking the observed integrated luminosity to the disk radial extent. This trend is found in the simulated luminosity and shown in Fig. \ref{data_model}. For each disk mass bin, models with larger $R_{\rm c}$ (larger symbols) show higher $^{12}$CO and $^{13}$CO line luminosity than models with smaller $R_{\rm c}$ (smaller symbols), independent on the stellar properties. A similar trend was also found for more extended disk models (see Fig \ref{data_model_Miotello2016}), but the increase of luminosity due to larger critical radii was modest, as the emission was mostly optically thin, especially for disks with masses smaller than $10^{-3}M_{\odot}$. 

   \begin{figure*}
   \centering
   \label{12CO_13CO_fig}
   \includegraphics[width=13cm]{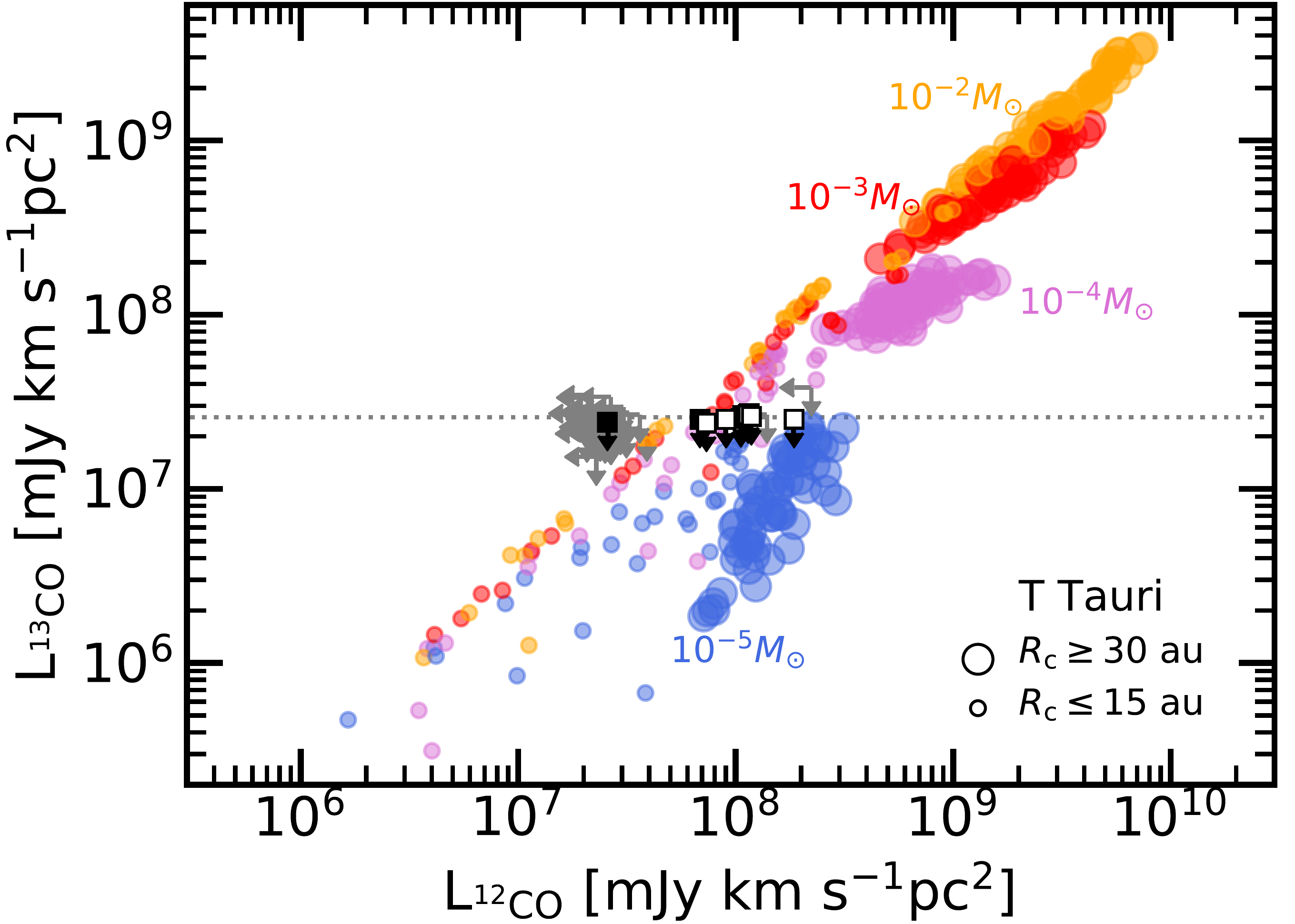}
      \caption{Simulated $^{12}$CO ($J=2-1$) versus $^{13}$CO ($J=2-1$) line luminosity of T Tauri disk models are presented: disk masses are color coded and different values of the critical radius $R_{\rm c}$ are shown by different symbol sizes. DALI model results from this work ($R_{\rm c}$= 0.5, 1, 2, 5, 15 au) are shown by smaller circles, while results from \cite{Miotello2016} ($R_{\rm c}$= 30, 60, 200 au) are shown by larger symbols. Lupus $^{12}$CO and $^{13}$CO ($J=2-1$) line luminosity are presented with black squares (empty squares if the $^{12}$CO line is partially absorbed by the cloud), where the $^{13}$CO non detections are shown as upper limits by the black arrows and the $^{12}$CO (and $^{13}$CO) non detections are shown as upper limits by the gray arrows. The dotted gray line shows the average of the $^{13}$CO Lupus upper limits.
      }
   \end{figure*}

Finally, $^{12}$CO and $^{13}$CO ($J=2-1$) integrated line luminosities are shown for compact ($R_{\rm c}\leq 15$ au, this work) and extended \citep[$R_{\rm c}\geq 30$ au,][]{Miotello2016} disk models in Fig. \ref{12CO_13CO_fig}. The two sets of models cover different regions of the luminosity-luminosity space, with more extended disks models resulting in higher $^{12}$CO and $^{13}$CO integrated line luminosities. The dotted grey line in Fig. \ref{12CO_13CO_fig} shows the median of the $^{13}$CO upper limits \citep{Ansdell2018}. As discussed earlier, the simulated $^{13}$CO and $^{12}$CO emission obtained with compact disk models is optically thick. Similar conditions are found also for very massive extended disk models, i.e. with disk masses larger than $10^{-3}M_{\odot}$. The extended disk models with $M_{\rm disk}=10^{-2}M_{\odot}$, shown with the large yellow symbols, produce $^{12}$CO and $^{13}$CO integrated line luminosity which qualitatively follow the same trend of the compact disk model results (small symbols). On the other hand, $^{13}$CO emission of less massive extended disk models deviate from the optically thick regime, bending to a steeper trend as $^{13}$CO optically thinner emission scales directly with mass. 

When comparing the sample of faint Lupus disks considered in this work and the model results presented in Fig. \ref{12CO_13CO_fig}, it is clear that by improving in sensitivity, i.e. by re-observing the faintest disks which were not detected in $^{12}$CO and $^{13}$CO for longer integration times, it would be possible to discriminate between the two scenarios: low-mass extended disks versus compact disks. 
This would be especially interesting for the disks in group A, as it is not possible to constrain from current observations if they are compact or extended but with very low CO surface density. 

\section{Discussion}
\label{sec:discussion}
The compact disk models presented in this paper add on the large grid of extended models published by \cite{Miotello2016}, filling a new part of the parameter space not extensively sampled by DALI models before. As shown in Fig. \ref{data_model}, such new models are consistent with ALMA observations of the fainter disks in the Lupus SFR \citep{Ansdell2018} and could provide a simple solution to the problem of faint CO isotopologue emission in disks. 

Faint CO isotopologue observations of disks have been recently interpreted as a sign of quick chemical evolution. This hypothesis is supported by \emph{Herschel}-PACS observations of the HD fundamental line in few bright disks. These observations showed that CO-based gas masses can be order(s) of magnitude smaller than HD-based disk masses \citep[e.g.,][]{Bergin2013,Favre2013}. This potential inconsistency has been explained by locking up of volatiles as ice in larger bodies, leading to low observed CO fluxes and this is supported by observations and modelling of other molecular species such as C$_2$H and N$_2$H$^+$ \citep[see e.g.,][and references therein]{Cleeves2018,Miotello2019,Anderson2019,Fedele2020}. It is still not clear if such a scenario, that was tested uniquely for bright and extended disks, applies also in the case of compact disks and it is in principle not in conflict with the results presented in Sec. \ref{Sec:results}. However, chemical models need the presence of an icy midplane in order to efficiently lock C and O in less volatile species \citep[see e.g.,][]{Eistrup2016,Eistrup2018,Bosman2017,Bosman2018}. Compact disk models are however generally warmer and the reservoir of frozen molecular material is reduced, compared to extended disk models (see Fig \ref{2D}). If such compact disks exist, their CO isotopologue emission may therefore simply be faint because of their reduced radial size, while volatile C and O depletion may be the main factor reducing CO fluxes in more extended and colder disks. 

Comparing our compact disk models with observations in Lupus allows us to constrain the fraction of disks that are potentially compact and optically thick. The entire Lupus sample studied by \cite{Ansdell2016,Ansdell2018} is composed of 99 disks, out of which 11 are in the group A and 52 in the group B, shown in Fig. \ref{data_model_Miotello2016}. Disks in group B, which are the 51.5\% of the sample, are incompatible with extended disk models, unless C and O are largely depleted. Disks in group A, are in principle compatible with both extended and faint disks, or compact and thick disks. Potentially the fraction of compact disks in Lupus could be up to 62.4\%, if we also consider disks in group A. To date there are no available ALMA observations for a sample of faint Class II disks that are deep enough to discriminate between the two scenarios for the sources in group A, as mainly the brightest end of disks observed by the ALMA disk surveys have been followed up at higher sensitivity and angular resolution. One order of magnitude deeper $^{12}$CO and $^{13}$CO observations of faint disks at a moderate angular resolution of $0.1-0.3"$, i.e., reaching integrated line luminosities $\sim 2 \times 10^{6}$ mJy km s$^{-1}$ pc$^{2}$ (see Fig. 3)\footnote{This luminosity is set by the minimum simulated $^{13}$CO integrated flux of the $10^{-5} M_{\odot}$ extended disk models, i.e., 3.72$\times 10^{-3}$ K km/s \citep[see][for more detail]{Miotello2016}}, will give us the opportunity to discriminate between two scenarios: very compact unresolved disks ($R_{\rm c}\lesssim 15$ au) whose emission is optically thick versus extended disks, whose faint optically thin emission is due to their low mass. If the sensitivity is improved by one order of magnitude, Lupus disks in group A that are already detected in $^{12}$CO will be most likely detected in $^{13}$CO, and will be either compatible with compact or extended disk models. In the latter case, their CO emission should also be resolved, which should instead not be the case if they are compatible with compact disk models. Either way, such observations would be a valuable test to our physical-chemical disk models. The potential of this approach is based on the fact that, for compact disk models, CO emission is optically thick, and the integrated flux scales with the disk size. Therefore, no high-resolution observations are needed, as for compact disks one would not need to resolve the CO emission to constrain their radial extent. 

If the faint end of the Lupus disks population were due to very compact disks, this would challenge viscous evolution theory which would predict extended gaseous disks. 
\cite{Trapman2020}, for example, have managed to reproduced the $^{12}$CO fluxes of the bright Lupus disks with viscously evolving disk models. In their Fig 6 they show viscous disks with initial conditions that are tuned to reproduce the average mass accretion rate in Lupus, that have observed sizes of at least $\sim$ 100 au, much more extended than what is predicted by our compact disk models. Furthermore, to consider more broadly the effect of initial conditions, in the regime of fast viscous spreading at time $t$ the viscous time $t_{\nu}$ is such that $t_{\nu}(R_{\rm c}) \sim t$ \citep{Lynden-BellPringle1974,Hartmann1998}. This relation is a lower limit on $R_{\rm c}$, because the disk could be born with an initially large size and therefore be slowly spreading. With $\alpha_{\rm visc}\sim10^{-3}$ and $t\sim$2 Myr, this implies a $R_{\rm c}$ of at least $\sim$40 au. Therefore,  $R_{\rm c} \lesssim $ 15 au, as in our compact disk models, implies low values of $\alpha_{\rm visc} \lesssim 4 \times 10^{-4}$. This is at the lower end of the range of values predicted by the magneto-rotational instability. Therefore $R_{\rm c} \lesssim $ 15 au would set strong constraints on the amount of viscosity and cast doubts on whether accretion is driven by viscosity rather than by an alternative mechanism such as MHD disc winds. Other processes should therefore be invoked to truncate disks to such small sizes, as for example external photoevaporation or the encounter with another star. However, we do not expect any of these processes to be relevant in a Star-forming region such as Lupus \citep[see e.g.,][]{Winter2018}. An interesting  implication to planet formation is that, in such small and optically thick disks, there may be substantial reservoirs of gas for forming Jupiter-like planets within Jupiter's orbital radius.

Irrespective of the physical interpretation, the compact disk models presented here offer a new instrument for the interpretation of current and future ALMA observations, among other examples, binary disks, which extend to smaller outer radii \citep[][Rota et al., in prep.]{Manara2019,Zurlo2020}. The simulated integrated continuum and CO isotopologue luminosities are reported in Appendix \ref{ancillary}.

\section{Conclusions}
Results from a new grid of compact disk models, with critical radius $R_{\rm c}=0.5,1,2,5,15$ au, run with DALI are presented in this paper. Such model results are consistent with $^{12}$CO and $^{13}$CO fluxes of the fainter Lupus disks, which could be explained by more extended disk models only if volatile C and O were largely depleted by orders of magnitude. The main conclusions from this work are the following:
   \begin{enumerate}
      \item Lupus disks that are not detected in $^{13}$CO emission, and with faint or undetected $^{12}$CO emission, may be intrinsically compact. The fraction of compact disks is potentially between roughly 50\% and 60\% of the entire Lupus sample;
      \item One order of magnitude deeper $^{12}$CO and $^{13}$CO observations - compared with ALMA disk surveys observations - of faint disks at an angular resolution of $0.1-0.2"$ will give us the opportunity to discriminate between two scenarios: very compact unresolved disks ($R_{\rm c}\lesssim 15$ au) whose emission is optically thick versus extended and resolved disks, whose faint emission is optically thin;
      \item The simulated integrated continuum and CO isotopologue fluxes of the new grid of compact models are reported in Appendix \ref{ancillary}.
   \end{enumerate}

\begin{acknowledgements}
      The authors wish to thank the anonymous referee for insightful comments and Ewine van Dishoeck, Leonardo Testi, Ted Bergin, and Antonella Natta for useful discussions. This project has received funding from the European Unions Horizon 2020 research and innovation programme under the Marie Sklodowska-Curie grant agreement No 823823, (RISE DUSTBUSTERS). This work was funded by the Deutsche Forschungsgemeinschaft (DFG, German Research Foundation) – Ref no. FOR 2634/1 ER685/11-1. G.R. acknowledges support from the Netherlands Organisation for Scientific Research (NWO, program number 016.Veni.192.233) and from an STFC Ernest Rutherford Fellowship (grant number ST/T003855/1). 
\end{acknowledgements}

%
%

\bibliographystyle{aa}
\bibliography{bibliography}

\begin{appendix} 

\section{Line intensities and disk masses}
\label{L-mass}
Similarly to what done by \cite{Miotello2016} and \cite{Miotello2017},  it is possible to determine how line intensities depend on the
disk mass by computing the medians of the $^{13}$CO and C$^{18}$O $(J=2-1)$ line intensities obtained by the compact disk models\footnote{Only for $i=10^{\rm o}$. A similar exercise can be done for other inclination angles using the model results in Table B.2.} (see disk parameters in Table \ref{table:1}) in different disk mass bins.

   \begin{figure}[h]
   \centering
   \includegraphics[width=9cm]{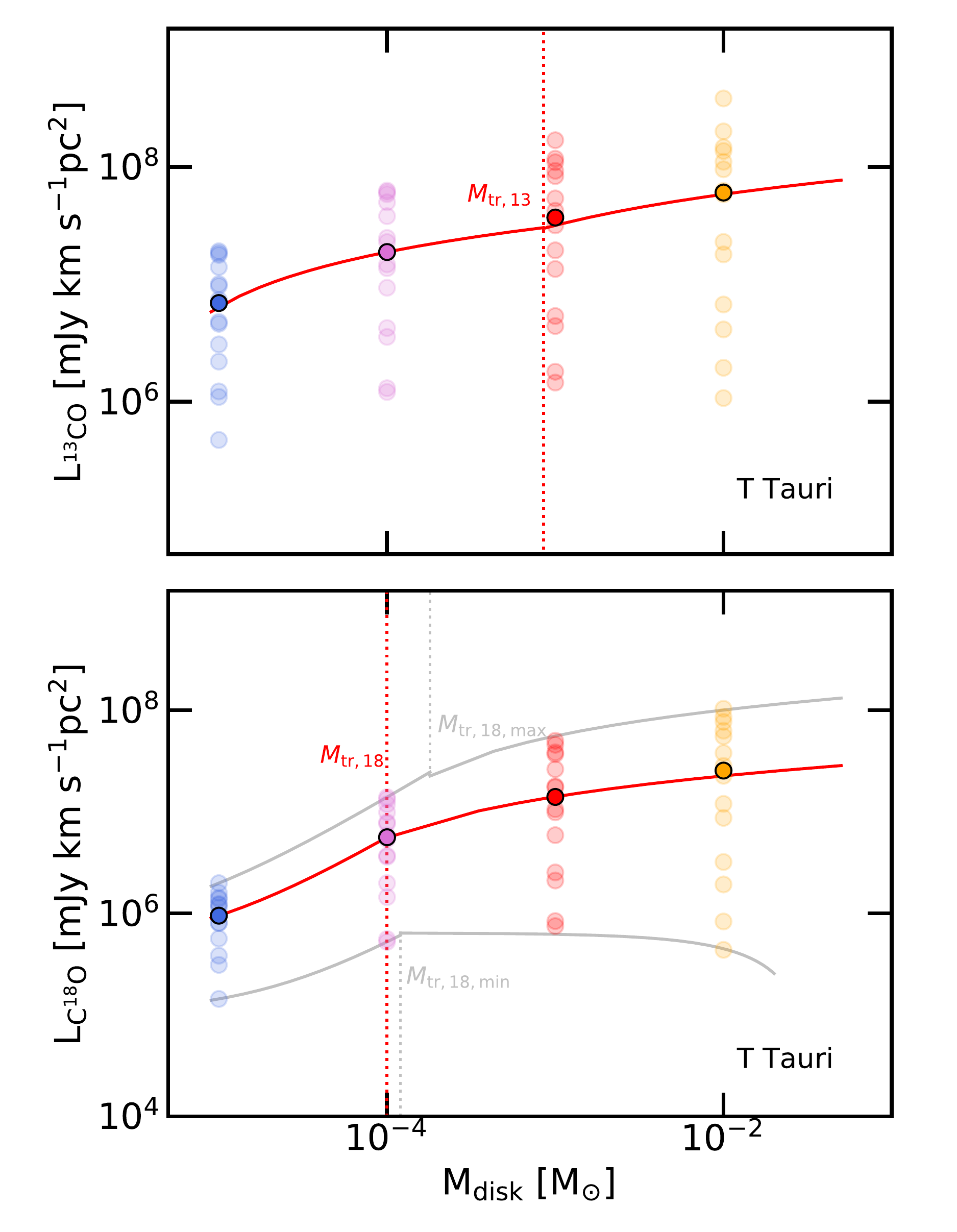}
      \caption{Median of the $^{13}$CO (upper panel) and C$^{18}$O (lower panel) $J=2-1$ line luminosities in different mass bins for compact disk models are presented with filled circles. Red solid lines show the fit functions presented in Eq. \ref{eq_13} and \ref{eq_18}.  The translucent symbols show all model results, while the grey solid lines show the fit to the maximum and minimum luminosities, that can be used to estimate uncertainties on the mass measurements.}
         \label{CO_mass}
   \end{figure}

These trends are presented in Fig. \ref{CO_mass}, both for $^{13}$CO (upper panel) and C$^{18}$O (lower panel). Being optically thick, $^{13}$CO intensity does not increase linearly with mass, but can be fitted to a logarithmic function of the disk mass. On the contrary, C$^{18}$O emission is optically thin for disk masses smaller than $M_{\rm tr}=10^{-4}M_{\odot}$, and it scales linearly with mass. The $^{13}$CO and C$^{18}$O $(J=2-1)$ line luminosities can be expressed by the following fit functions of the disk mass:
\begin{equation}
L_{\rm ^{13}CO} =
  \begin{cases}
    6.8\times10^{7} + 5.4\times10^{7} \, \text{log}_{10}(M_{\rm disk})       & \quad \text{if } M_{\rm disk}\leq M_{\rm tr,13}\\
    1.1\times10^{8} + 1.2\times10^{7} \, \text{log}_{10}(M_{\rm disk})       & \quad \text{if } M_{\rm disk}>M_{\rm tr,13}
  \end{cases}
\label{eq_13}
\end{equation}
and
\begin{equation}
L_{\rm C^{18}O} =
  \begin{cases}
    4.3\times10^{5} + 5.2\times10^{10}M_{\rm disk}       & \quad \text{if } M_{\rm disk}\leq M_{\rm tr,18}\\
    3.9\times10^{7} + 3.7\times10^{6} \, \text{log}_{10}(M_{\rm disk})       & \quad \text{if } M_{\rm disk}>M_{\rm tr,18}
  \end{cases}
\label{eq_18}
\end{equation}
where $M_{\rm tr,13}=8.5\times 10^{-4} M_{\odot}$ and $M_{\rm tr,18}=10^{-4} M_{\odot}$.

Since $^{13}$CO emission is always optically thick in compact disk models, only C$^{18}$O emission may be used as mass tracer, only if $M_{\rm disk}<M_{\rm tr,18}$, and assuming that no volatile C and O depletion is there. The fit to the maximum and minimum simulated luminosities, shown by the grey lines in Fig. \ref{CO_mass}, can be used to estimated uncertainties on the mass determinations, and they are as follows:
\begin{equation}
L_{\rm C^{18}O,max} =
  \begin{cases}
    6.5\times10^{5} + 1.3\times10^{11}M_{\rm disk}       & \quad \text{if } M_{\rm disk}\leq M_{\rm tr,18,max}\\
    1.9\times10^{8} + 1.9\times10^{7} \, \text{log}_{10}(M_{\rm disk})       & \quad \text{if } M_{\rm disk}>M_{\rm tr,18,max}
  \end{cases}
\label{eq_max}
\end{equation}
and
\begin{equation}
L_{\rm C^{18}O,min} =
  \begin{cases}
    10^{5} + 4.2\times10^{9}M_{\rm disk}       & \quad \text{if } M_{\rm disk}\leq M_{\rm tr,18,min}\\
    6.4\times10^{5} - 1.9\times10^{7} M_{\rm disk}      & \quad \text{if } M_{\rm disk}>M_{\rm tr,18,min}
  \end{cases}
\label{eq_max}
\end{equation}
where $M_{\rm tr,18,max}=1.8\times 10^{-4}$ and $M_{\rm tr,18,min}=1.2\times 10^{-4}$.

\section{Ancillary material}
\label{ancillary}

The integrated $^{12}$CO $(J=2-1)$ fluxes for the compact disks in Lupus studied in this work are reported in Tab. \ref{table:Lupus}. These fluxes are measured using an aperture synthesis method, as done by \cite{Ansdell2018}. For non detections, the 3$-\sigma$ upper limits are reported. Cloud absorption may be affecting a few of the Lupus sources considered in this work \citep[see $^{12}$CO spectra in Fig. 11 of][]{Ansdell2018}, and integrated $^{12}$CO luminosities should be considered as meaningful lower limits. In fact, we do not expect such absorption to reduce the effective disk $^{12}$CO emission more than a factor of two.

\begin{table*}[h]
\caption{$^{12}$CO and $^{13}$CO fluxes and 3$\sigma$ upper limits, distances \citep{GaiaDR2}, stellar luminosity, and stellar mass - calculated using the evolutionary tracks by \cite{Baraffe2015} \citep{Alcala2017} - of the sources in the studied sub-sample of faint Lupus disks.} \label{table:Lupus}      
\centering                          
\begin{tabular}{rlllll}        
\hline\hline                 
Source & $F_{\rm ^{12}CO}$ & $F_{\rm ^{13}CO}$ & $d$ & $L_{\star}$ & $M_{\star}$\\   
& [mJy km s$^{-1}$] & [mJy km s$^{-1}$] & [pc] & $L_{\odot}$ & $M_{\odot}$ \\
\hline
              Sz77 &  350.0 $\pm$ 80.0 & $<$ 69.0 & 155 & 0.59 & 1.09\\      
             Sz114 &   $<$ 85.2 & $<$ 87.0 & 162 & 0.21 & 0.21\\         
 J16100133-3906449 &   $<$ 84.0 & $<$ 90.0 & 193 & 0.19 & -\\           
 J16080175-3912316 &   $<$ 81.9 & $<$ 87.0 & 159$^{b}$ & - & -\\           
 J16121120-3832197 &   $<$ 87.3 & $<$ 90.0 & 159$^{b}$ & - & -\\           
              Sz90 &  577.0 $\pm$ 88.0$^{a}$ & $<$ 87.0 & 160 & 0.42 & 1.03\\     
 J16104536-3854547 &   $<$ 87.0 & $<$ 87.0 & 159$^{b}$ & - & -\\           
 J16134410-3736462 &   $<$ 85.5 & $<$ 84.0 & 159 & 0.04 & 0.15\\    
             Sz88A &   $<$ 83.4 & $<$ 84.0 & 158 & 0.31 & 0.87\\
             Sz106 &   $<$ 86.7 & $<$ 90.0 & 162 & 0.06 & 0.59\\
 J16085324-3914401 &  322.0 $\pm$ 67.0$^{a}$ & $<$ 87.0 & 168 & 0.21 & 0.38\\   
 J16101857-3836125 &   $<$ 85.5 & $<$ 90.0 & 159$^{b}$ & 0.04 & 0.14\\ 
           Sz123A  &   $<$ 90.0 & $<$ 90.0 & 159 & 0.13 & 0.65\\ 
 J16002612-4153553 &   $<$ 84.6 & $<$ 90.0 & 164 & 0.08 & 0.14\\ 
              Sz95 &   $<$ 86.1 & $<$ 87.0 & 158 & 0.26 & 0.40\\ 
              Sz96 &  221.0 $\pm$ 55.0 & $<$ 87.0 & 157 & 0.42 & 0.75 \\
             Sz117 &   $<$ 85.2 & $<$ 87.0 & 159 & 0.28 & 0.35\\
 J16000060-4221567 &   $<$ 85.2 & $<$ 87.0 & 161 & 0.10 & 0.22\\
             Sz131 &  367.0 $\pm$ 57.0$^{a}$ & $<$ 90.0 & 160 & 0.15 & 0.36\\
              Sz72 &  240.0 $\pm$ 31.0$^{a}$ & $<$ 90.0 & 156 & 0.27 & 0.54\\
            Sz123B &   $<$ 89.1 & $<$ 84.0 & 159 & 0.03 & 0.45\\
              Sz66 &  373.0 $\pm$ 84.0 & $<$ 87.0 & 157 & 0.22 & 0.38\\
 J16085529-3848481 &   $<$ 87.0 & $<$ 87.0 & 158 & 0.05 & 0.08\\    
              Sz74 &   $<$ 66.0 & $<$ 87.0 & 159 & 1.16 & -\\    
 J16101307-3846165 &   $<$ 82.8 & $<$ 87.0 & 159$^{b}$ & - & -\\    
              Sz97 &   $<$ 83.7 & $<$ 87.0 & 157 & 0.11 & 0.25 \\    
              Sz99 &   $<$ 87.3 & $<$ 87.0 & 159 & 0.05 & 0.23\\    
             Sz103 &   $<$ 84.3 & $<$ 90.0 & 160 & 0.12 & 0.26\\    
 J16084940-3905393 &   $<$ 86.1 & $<$ 87.0 & 159 & - & -\\    
             Sz110 &   $<$ 88.5 & $<$ 93.0 & 159 & 0.18 & 0.28\\    
             Sz113 &   $<$ 84.3 & $<$ 90.0 & 163 & 0.04 & 0.17\\    
             Sz115 &   $<$ 85.8 & $<$ 90.0 & 158 & 0.11 & 0.22\\    
             Sz88B &   $<$ 84.3 & $<$ 87.0 & 159 & 0.07 & 0.21\\    
 J15592523-4235066 &   $<$ 87.0 & $<$ 84.0 & 159$^{b}$ & 0.02 & 0.12\\    
 J16081497-3857145 &  343.0 $\pm$ 56.0 & $<$ 87.0 & 158 & 0.01 & 0.09\\
 J15445789-3423392 &   $<$ 85.5 & $<$ 90.0 & 159$^{b}$ & - & - \\
             Sz104 &   $<$ 84.3 & $<$ 87.0 & 166 & 0.07 & 0.17\\
             Sz112 &   $<$ 84.3 & $<$ 90.0 & 160 & 0.12 & 0.18 \\
 J16115979-3823383 &   $<$ 87.3 & $<$ 84.0 & 159$^{b}$ & - & - \\ 
 J16073773-3921388 &   $<$ 86.1 & $<$ 90.0 & 174 & - & - \\
 J16080017-3902595 &   $<$ 85.8 & $<$ 87.0 & 159 & - & - \\
  J160828.1-391310 &   $<$ 86.4 & $<$ 87.0 & 159$^{b}$ & - & - \\   
            Sz108B &   $<$ 86.1 & $<$ 87.0 & 169 & 0.11 & 0.1\\
 J16085373-3914367 &   $<$ 88.5 & $<$ 84.0 & 159 & - & 0.10\\
             Sz81A &   $<$ 89.2 & $<$ 87.0 & 160 & 0.25 & 0.26\\
 J16101984-3836065 &   $<$ 85.8 & $<$ 87.0 & 158 & 0.04 & 0.08\\
 J16095628-3859518 &   83.0 $\pm$ 25.0 & $<$ 90.0 & 157 & - & - \\
 J15430131-3409153 &   $<$ 207.0& $<$ 87.0 & 159$^{b}$ & - & - \\
 J15430227-3444059 &   $<$ 85.5 & $<$ 87.0 & 159$^{b}$ & - & - \\
 J15450634-3417378 &  303.0 $\pm$ 64.0$^{a}$ & $<$ 87.0 & 154 & - & -\\
 J16075475-3915446 &   $<$ 84.6 & $<$ 90.0 & 153 & - & - \\  
  J160831.1-385600 &   $<$ 86.4 & $<$ 87.0 & 159$^{b}$ & - & -\\
 J16085828-3907355 &   $<$ 87.0 & $<$ 90.0 & 159$^{b}$ & - & -\\
 J16085834-3907491 &   $<$ 88.5 & $<$ 87.0 & 159$^{b}$ & - & -\\
 J16091644-3904438 &   $<$ 92.4 & $<$ 87.0 & 159$^{b}$ & - & -\\
  J160918.1-390453 &   $<$ 136.5& $<$ 84.0 & 159$^{b}$ & - & -\\
 J16092032-3904015 &   $<$ 28.7 & $<$ 90.0 & 159$^{b}$ & - & -\\
 J16092317-3904074 &   $<$ 86.1 & $<$ 90.0 & 159$^{b}$ & - & -\\
  \hline
    \end{tabular}
  \end{table*}
\begin{table*}[h]
\centering 
\begin{tabular}{rlllll} 
\hline\hline    
Source & $F_{\rm ^{12}CO}$ & $F_{\rm ^{13}CO}$ & $d$& $L_{\star}$ & $M_{\star}$ \\   
& [mJy km s$^{-1}$] & [mJy km s$^{-1}$] & [pc] & $L_{\odot}$ & $M_{\odot}$ \\
\hline
  J160934.2-391513 &   $<$ 86.7 & $<$ 90.0 & 159$^{b}$ & - & -\\
 J16093928-3904316 &   $<$ 85.2 & $<$ 90.0 & 159$^{b}$ & - & -\\
 J16102741-3902299 &   $<$ 87.6 & $<$ 87.0 & 159$^{b}$ & - & -\\
 J16120445-3809589 &   $<$ 85.8 & $<$ 87.0 & 159$^{b}$ & - & -\\
          V856Sco &   $<$ 85.2 & $<$ 87.0 & 159$^{b}$ & - & -\\
\hline   

\end{tabular}
\\
{\tiny $^{a}$: By visual inspection of the spectra published by \cite{Ansdell2018}, $^{12}$CO line\\ seems partially absorbed by the cloud. $^{b}$: No distance has been measured by Gaia DR2\\ \citep{GaiaDR2}. The distance is therefore calculated as the median\\ of the Gaia DR2 distances measured for the other sources in this Lupus subsample.}
\end{table*}

The simulated spatially integrated continuum fluxes (at 880 $\mu$m in Jy) and of the CO isotopologues lines (in K km s$^{-1}$) for the new grid of compact T Tauri-like and VLMS-like disk models are reported in Tab.  \ref{longtable} and Tab. \ref{longtable_VLMS} respectively.

\begin{figure*}[h]
   \centering
   \includegraphics[width=16cm]{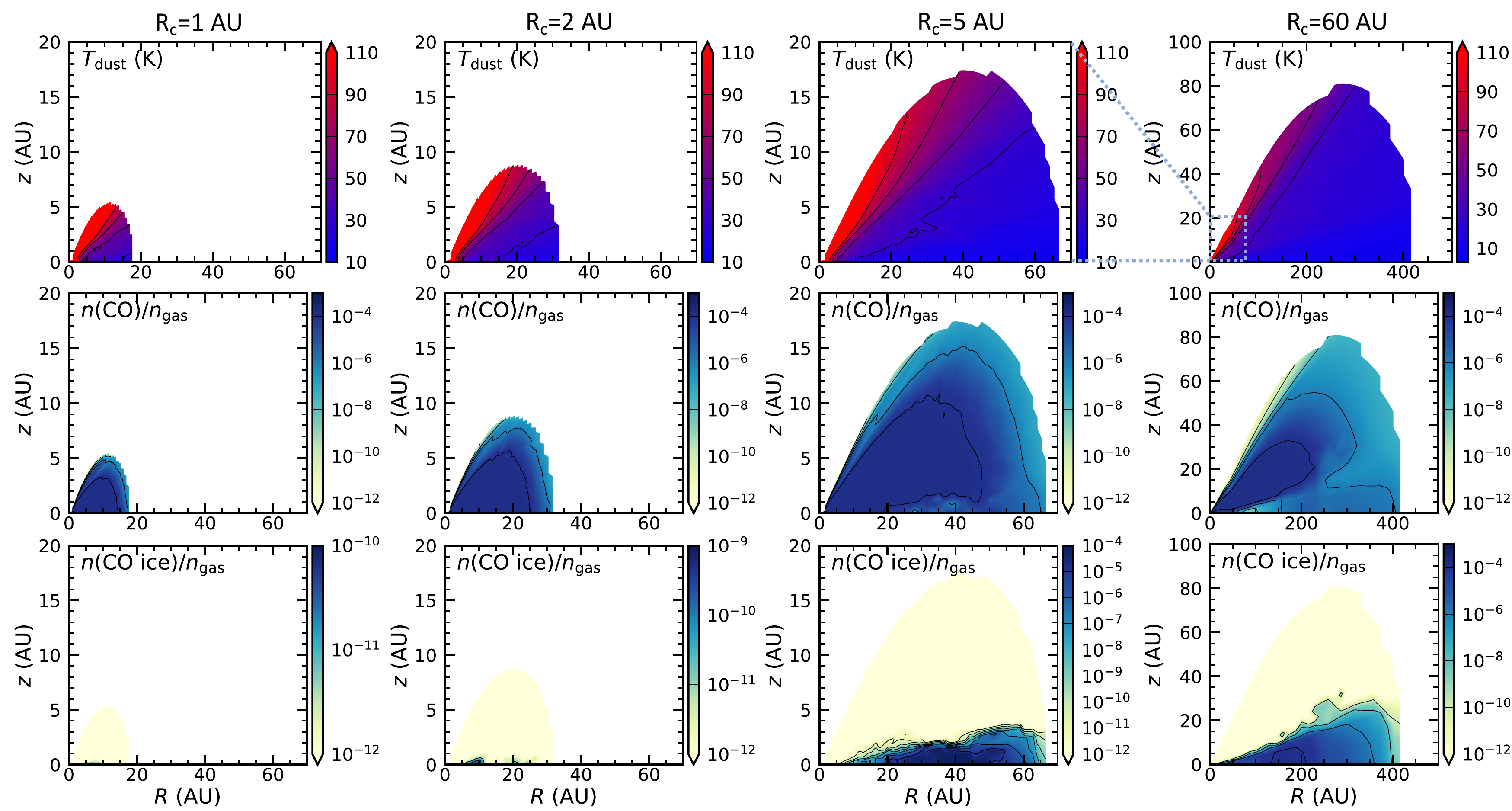}
      \caption{2D plots of the dust temperature structure, as well as the gaseous and ice CO abundance distribution for a selection of the compact T Tauri disk models: $M_{\rm disk}=10^{-3}M_{\odot}$, $\gamma=1$, and $R_{\rm c}= 1$ au (left), 2 au (middle left), 5 au (middle right), 60 au (right). Note the change of scale in the $R$, and $Z$ axes in the right panels.}
         \label{2D}
   \end{figure*}

The 2D plots of the dust temperature structure, as well as the gaseous and ice CO abundance distribution for a selection of the compact disk models are shown in Fig. \ref{2D}. For a fixed disk mass, more compact disks are warmer than more extended disks, with $T_{\rm dust}$ higher than 20 K almost everywhere in the disk (if $R_{\rm c}\leq 2$ au). As a consequence the amount of CO frozen onto grains is negligible.

\begin{figure}[h]
   \centering
   \includegraphics[width=9cm]{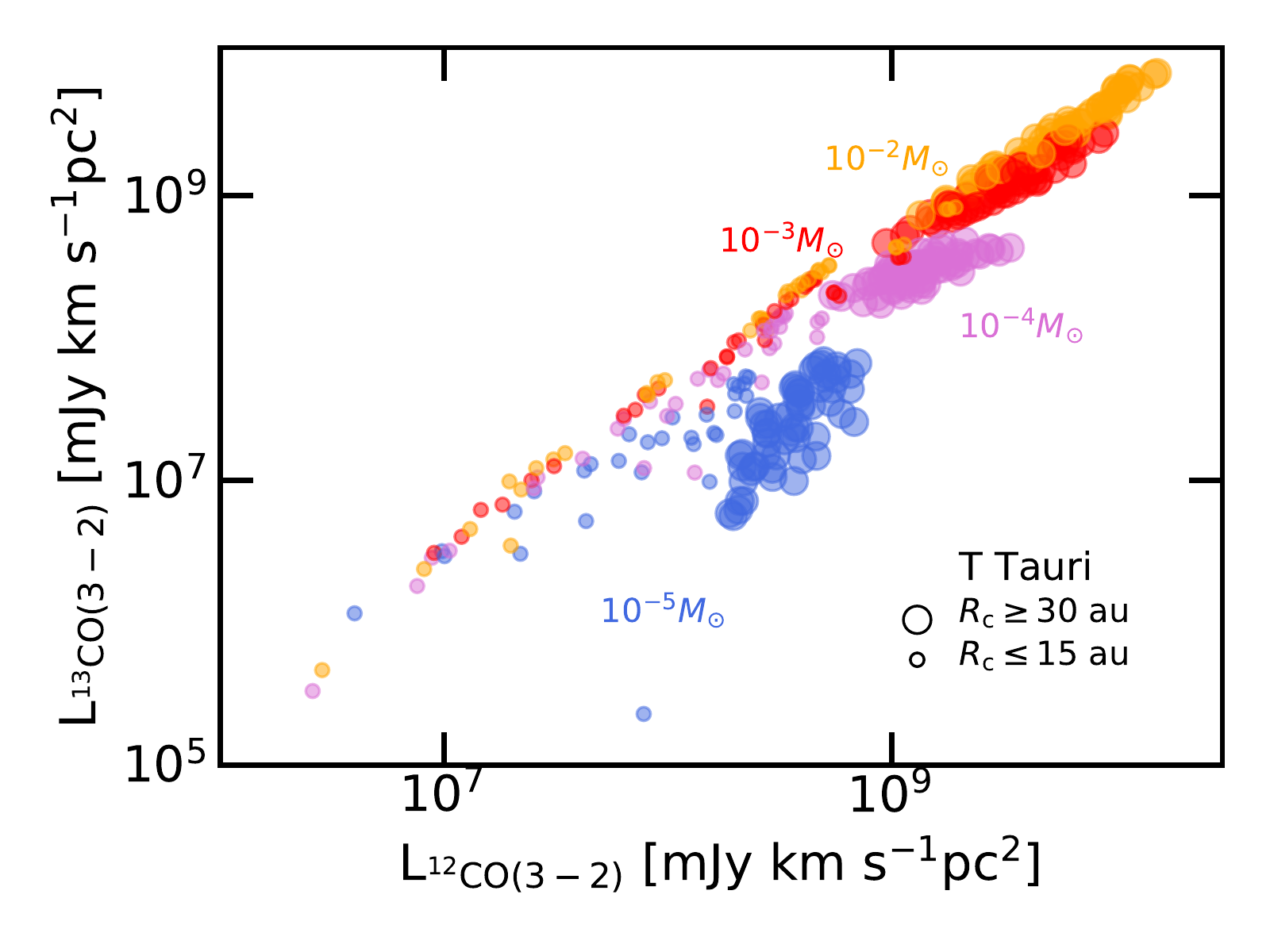}
      \caption{Simulated $^{12}$CO $(J=3-2)$ versus $^{13}$CO $(J=3-2)$ line luminosity of T Tauri disk models are presented: disk masses are color coded and different values of the critical radius $R_{\rm c}$ are shown by different symbol sizes. DALI model results from this work $(R_{\rm c}=0.5, 1, 2, 5, 15$ au) are shown by smaller circles, while results from \cite{Miotello2016} ($R_{\rm c}=30, 60, 200$ au) are shown by larger symbols. }
         \label{12CO_13CO_32}
   \end{figure}

Simulated $^{12}$CO $(J=3-2)$ versus $^{13}$CO $(J=3-2)$ line luminosity of T Tauri disk models are presented in Fig. \ref{12CO_13CO_32}: disk masses are color coded and different values of the critical radius $R_{\rm c}$ are shown by different symbol sizes. DALI model results from this work $(R_{\rm c}=0.5, 1, 2, 5, 15$ au) are show by smaller circles, while results from \cite{Miotello2016} ($R_{\rm c}=30, 60, 200$ au) are shown by larger symbols.

\begin{table*}[h]
\caption{Disk parameters and integrated fluxes simulated with the grid of compact T Tauri-like models. The line ray-tracing has been computed assuming a distance of 150 pc.}
\label{longtable}      
\centering                          
\begin{tabular}{ccccccccccc}       
\hline
\hline
$R_{\rm c}$	& $\gamma$	& $M_{\rm disk}$	& $i$	& $F_{\rm cont}$	& $F_{\rm ^{12}CO (2-1)}$ & $F_{\rm ^{13}CO (2-1)}$ & $F_{\rm C^{18}O (2-1)}$	& $F_{\rm ^{12}CO (3-2)}$ & $F_{\rm ^{13}CO (3-2)}$ & $F_{\rm C^{18}O (3-2)}$\\
		 (au)	&	& ($M_{\odot}$)	& (deg)	& (Jy)	& (K km s$^{-1}$)	& (K km s$^{-1}$)	& (K km s$^{-1}$) & (K km s$^{-1}$)	& (K km s$^{-1}$)	& (K km s$^{-1}$)\\
\hline 
1.0	&	1.0	 &	1.e-5	& 10	& 1.974038e-04 &   	7.106702e-03  	& 1.952710e-03	 & 5.058933e-04 & ... & ... & ...\\
1.0	&	1.0	 &	1.e-4	& 10	& 9.544535e-04 &   	9.047580e-03  	& 3.167431e-03	 & 1.293363e-03 & ... & ... & ...\\
1.0	&	1.0	 &	1.e-3	& 10	& 1.557710e-03 & ... & ... & ... & ... & ... & ... \\
1.0	&	1.0	 &	1.e-2 & ...	& ... & ... & ... & ...& ... & ... & ... \\
\hline                                   
\end{tabular}
\end{table*}

\begin{table*}[h]
\caption{Disk parameters and integrated fluxes simulated with the grid of compact VLMS-like models. The line ray-tracing has been computed assuming a distance of 150 pc.}
\label{longtable_VLMS}      
\centering                          
\begin{tabular}{ccccccccccc}       
\hline
\hline
$R_{\rm c}$	& $\gamma$	& $M_{\rm disk}$	& $i$	& $F_{\rm cont}$	& $F_{\rm ^{12}CO (2-1)}$ & $F_{\rm ^{13}CO (2-1)}$ & $F_{\rm C^{18}O (2-1)}$	& $F_{\rm ^{12}CO (3-2)}$ & $F_{\rm ^{13}CO (3-2)}$ & $F_{\rm C^{18}O (3-2)}$\\
		 (au)	&	& ($M_{\odot}$)	& (deg)	& (Jy)	& (K km s$^{-1}$)	& (K km s$^{-1}$)	& (K km s$^{-1}$) & (K km s$^{-1}$)	& (K km s$^{-1}$)	& (K km s$^{-1}$)\\
\hline 
1.0	&	1.0	 &	1.e-5	& 10	& 1.024040e-04 &	1.518841e-03 &	4.120504e-04 &	5.221508e-03 & ... & ... & ... \\
1.0	&	1.0	 &	1.e-4	& 10	& 4.297588e-04 &	2.523754e-03 &	1.009048e-03 &	6.596197e-03 & ... & ... & ... \\
1.0	&	1.0	 &	1.e-3	& 10	& 7.848836e-04 & ... & ... & ... & ... & ... & ... \\ 
1.0	&	1.0	 &	1.e-2 & ...	& ... & ... & ... & ... & ... & ... & ... \\
\hline  
\end{tabular}
\end{table*}

\end{appendix}

\end{document}